\documentclass[11pt,twoside]{article}

\usepackage{asp2006}
\usepackage{epsf}
\usepackage{psfig}
\usepackage{graphicx}
\usepackage{lscape}

\begin{document}
\title{BL Lac Objects in the Sloan Digital Sky Survey (SDSS)}   
%%% Fill in title
\author{A. C. Gupta$^{1}$, W. Yuan$^{1}$, X. Dong$^{2}$, T. Ji$^{2}$, H.-Y. Zhou$^{2}$ \& J. M. Bai$^{1}$}   
%%% Fill in author names
\affil{$^{1}$ National Astronomical Observatories/Yunnan Observatory, CAS, P.O. Box 110, Kunming, Yunnan
650011, China}    %%% Fill in author affiliations
\affil{$^{2}$ Center for Astrophysics, University of Science and Technology of China, Hefei, Anhui, China}
\begin{abstract} %%% Abstract to run on from here.
We collected a sample of 661 confirmed and 361 possible BL Lac candidates from the recent catalog of
BL Lac objects (Veron-Cetty \& Veron 2006). We searched these sources in the recent data release DR5 of
the Sloan Digital Sky Survey (SDSS) and found spectra were available for 169 and 109 confirmed and 
possible BL Lac candidates respectively. We found 32 candidates from confirmed and 19 candidates from 
possible BL Lac lists have non featureless spectra and are thus possibly not BL Lac candidates. We report 
here the preliminary results from our analysis of a sample of 278 BL Lac objects. 
\end{abstract}
%%% MAIN BODY OF TEXT GOES HERE. CONSULT "INSTRUCTIONS FOR AUTHORS USING
%%% LATEX2E MARKUP", SECTIONS 2.3-2.6 FOR HELP WITH EQUATIONS, FIGURES,
%%% AND TABLES.
\section{Introduction}   %%% Top level section head (remove "%" symbol)
A small subset of radio-loud active galactic nuclei (AGNs) show rapid variability
at complete EM spectrum with the emission being strongly polarized. Such AGNs are
called blazars and their radiation at all wavelength is predominantly non-thermal.
BL Lac objects are a subclass of blazars which have featureless spectrum, so it is
difficult to estimate the distances, luminosities and redshifts of these objects. 
In a small population of BL Lacs, host galaxies are also seen. A careful study of 
these sources can give few weak emission lines which will be helpful to determine 
the redshifts. In the present work based on the sample of 169 confirmed and 109
possible BL Lacs, our main aim to find genuine BL Lacs and separate out BL Lacs
and non BL Lacs in two different lists.    

\section{Optical Spectrum Analysis and Results}

We searched the spectra of 1022 BL Lacs sample in SDSS DR5 catalog and 
found spectra were available for 278 sources. The spectra of these sources were 
extracted and corrected for the galactic extinction using the extinction curve of
(Schlegel et al. 1998). Extinction corrected spectras were smoothed with a boxcar 
of 5 pixels for illustration. Then we visually inspected spectra of all sources 
and based on that we divided the spectra into two groups: (i) featureless spectrum 
(BL Lac objects), (ii) spectrum with emission and/or absorption line features (non BL 
Lacs). \\  
In our visual inspection, we found 137 sources from the list of confirmed and 90 sources
from possible BL Lacs have featureless spectrum i.e. confirmed BL Lacs. We found respectively 
32 and 19 sources from confirmed and possible candidate lists are non featureless and so 
classified as possibly non BL Lacs. Further analysis is going on.  

\begin{figure}
\center
\includegraphics[width=3.1in, height=3.1in, angle=0]{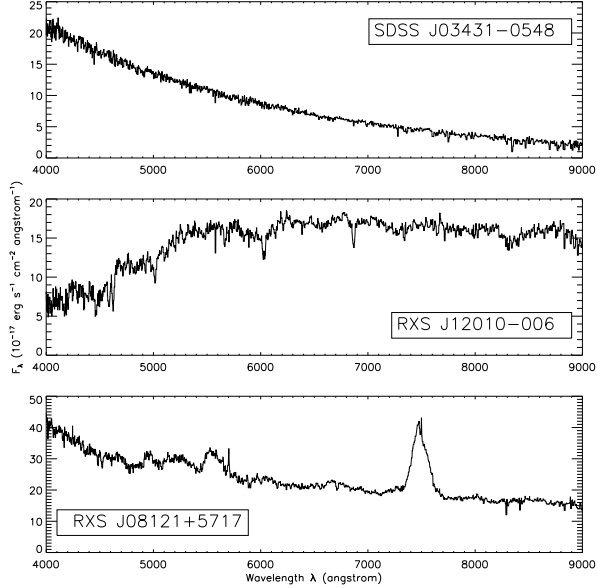}
\caption{Top, middle and bottom panels show the extinction corrected spectrum of confirmed 
BL Lac (form possible BL Lac list), non BL Lac (form possible BL Lac list) and non BL Lac 
(form confirmed BL Lac list) respectively.} 
\end{figure}

\noindent
\section{Conclusion \& Future Plan}

In the SDSS spectra of our sample of 278 BL Lac objects, we found 227 sources are genuine
BL lacs and 51 possibly belongs to non BL Lacs. We have following strategy for further detail
analysis of spectra to study these sources in more detail. \\
{\bf 1.} ~Sources having non featureless spectrum, redshift will be determined. \\
{\bf 2.} Using multi-wavelength published data of these sources, we will be able to get the 
spectral energy distribution (SED). \\  
{\bf 3.} We also have plan to search for new BL Lacs in SDSS DR5 data release. We will do proper
modeling of stellar component (Lu et al. 2006) and decompose AGNs spectra into stellar and non stellar
nuclear components, provided that two components are comparable in strength.  
%\subsection{}   %%% Second level section head (remove "%" symbol)
%\subsubsection{}   %%% Lowest level section head (remove "%" symbol)
%\section*{}    %%% Unnumbered top level section head (remove "%" symbol)
%\subsection*{}   %%% Unnumbered second level section head (remove "%" symbol)
\acknowledgements %%% Text of acknowledgements runs on after this command.
The work is supported by National Natural Science Foundation of China (NSF-10533050).

%%% THE BIBLIOGRAPHY
%%%
%%% CONSULT SECTION 3 OF "INSTRUCTIONS FOR AUTHORS" FOR HOW TO USE NATBIB.
%%% AUTHORS ARE ENCOURAGED TO USE EITHER THE "THEBIBLIOGRAPY" ENVIRONMENT
%%% BY UNCOMMENTING (DELETING THE "%" SYMBOL) THE COMMANDS BELOW, OR BY
%%% USING THE BIBTEX ENVIRONMENT. TO FIND OUT WHICH IS APPLICABLE TO YOUR
%%% CONTRIBUTION, CONSULT THE VOLUME EDITORS FOR YOUR PROCEEDINGS.
%%%

\end{document}